\documentstyle[osa,eqsecnum,manuscript]{revtex}

\begin{document}

\tightenlines
\title{Quantum noise in optical fibers I: \\
 stochastic equations}

\author{P. D. Drummond$^{1}$ and J. F. Corney$^{1,2}$ }

\address{$^{1}$Department of Physics, The University of Queensland, St. Lucia, QLD 4072, Australia\\
$^{2}$Department of Mathematical Modelling, Technical University of Denmark, DK-2800 Lyngby, Denmark}

\date{\today{}}

\maketitle
\begin{abstract}
\noindent We analyze the quantum dynamics of radiation propagating in a single mode optical
fiber with dispersion, nonlinearity, and 
Raman coupling to thermal phonons. We start from a fundamental
Hamiltonian that includes the principal known nonlinear effects and quantum
noise sources, including linear gain and loss. Both Markovian and frequency-dependent,
non-Markovian reservoirs are treated. This allows quantum Langevin equations
to be calculated, which have a classical form except for additional quantum
noise terms. In practical calculations, it is more useful to transform to Wigner
or +$P$ quasi-probability operator representations. These result in stochastic
equations that can be analyzed using perturbation theory or exact numerical
techniques. The results have applications to fiber optics communications, networking, and sensor technology. 
\end{abstract}


\section{Introduction}

The propagation of electromagnetic radiation through optical fibers is the central
paradigm of optical communications and sensor technology. It is also a novel
physical system, due to the materials processing of fused silica, that leads
to single-mode behaviour with extremely low losses. Over short distances (depending
on the pulse intensity) the well-known nonlinear Schr\"{o}dinger equation can
describe most optical fibers with great accuracy, and leads to soliton behaviour,
as well as to many other effects. Over longer distances, a number of reservoir
effects intervene, including attenuation, Raman scattering, and the use of amplifiers
and filters to compensate for losses. At the quantum level, both the original
nonlinearity and the additional couplings to reservoirs can lead to quantum
noise - which modifies the predictions of the classical nonlinear Schr\"{o}dinger
equation.

In this paper, we analyze the effects of quantum noise in fiber
optics. This extends and explains in more detail earlier theoretical work in
this area, which led to the first prediction\cite{s14,2} and measurement\cite{Solexp}
of intrinsic quantum noise effects in optical solitons. The theory given here
includes a detailed derivation of the relevant quantum Hamiltonian. We include
quantum noise effects due to nonlinearities, Raman reservoirs and Brillouin
scattering. The Raman/Brillouin noise is modeled using a multiple Lorentzian
fit to measured fluorescence data, in order to estimate the Raman gain coefficients.
Both gain and loss effects are included. This treatment is unified with theory
of gain/loss reservoirs, which was also predicted\cite{s35} and observed\cite{s53}
to have large effects on soliton propagation. All these reservoirs are treated
without using the Markovian approximation, in order to accurately treat the
frequency-dependent reservoirs found in practical applications.

The purpose of this work is to lay the foundations of practical methods for calculating and numerically simulating quantum effects in nonlinear optical fibers. These are significant
whenever quantum-limited behaviour is important in communications, sensing,
or measurement with optical fibers.

We introduce the basic quantum Hamiltonian for an optical fiber in section \ref{NSM}.
This gives a Heisenberg equation of motion which reduces to the nonlinear Schr\"{o}dinger
equation in the classical limit. The equation of motion is extended in section
\ref{RH} to include Raman and Brillouin effects, with gain and absorption processes
considered in section \ref{GA}. The complete Heisenberg equation in section
\ref{CHE} is the central result of this paper. Stochastic partial differential
equations can be derived from the quantum equation, using the phase-space methods
outlined in section \ref{PSM}. Applications of these methods to practical examples
are reserved for a following paper (QNII)\cite{DruCor99b}.

\section{Nonlinear Schr\"{o}dinger Model}

\label{NSM} The interaction between photons in a fiber is mediated through
the dielectric material constituting the fiber. The coupling to the dielectric
introduces frequency dependent and time delayed behaviour. The complete Hamiltonian
and its derivation have been given in the literature\cite{s14,s208,s21,s90,Drumhill};
we will only go over the salient points here. The starting point is a Lagrangian
that generates the classical Maxwell's equations for a one-dimensional dielectric
waveguide, and that gives a Hamiltonian corresponding to the dielectric energy\cite{s208}:
\begin{eqnarray}
H_{D}=\int dV\left[ \frac{1}{2\mu }|{\mathbf{B}}|^{2}+\int ^{t}_{t_{0}}{\mathbf{E}}(t')\cdot \dot{\mathbf{D}}(t')dt'\right] \, \, , & \label{Hamtn1} 
\end{eqnarray}
 where the electric field \( \mathbf{E}=(\mathbf{D}-\mathbf{P})/\epsilon _{0} \)
includes the polarization response of the dielectric to an incident electric
displacement \( \mathbf{D} \). The field variables are then quantized by introducing
equal-time commutators between the canonical coordinates \( \mathbf{D} \) and
\( \mathbf{B} \). We note that, of course, it is also possible to make other
choices of canonical momenta. This choice corresponds to a dipole-coupled\cite{PZL},
rather than minimal-coupled fundamental Lagrangian. While different Lagrangians
are canonically equivalent, the present choice - originally introduced\cite{Hillery}
by Hillery and Mlodinow in applications to dielectric theory - has the advantage
of comparative simplicity. The Lagrangian must produce both the correct energy\cite{Bloembergen}
and Maxwell's equations, otherwise the conjugate momenta will contain an arbitrary
scaling, leading to incorrect commutation relations\cite{Hillery,s208}.

\subsection{Fiber-optic Hamiltonian}

The optical fiber treated will be a single transverse mode fiber with dispersion
and nonlinearity. Since boundary effects are usually negligible in experiments,
it is useful to first take the infinite volume limit, which effectively replaces
a summation over wave-vectors with the corresponding integral. We will start
with a single polarization direction (i.e., a polarization preserving fiber).
The more general case is summarized elsewhere\cite{Drummond}, and will be treated in detail
subsequently.  The basic normally ordered, nonlinear Hamiltonian for the fiber in this case
is\cite{s208}: 
\begin{equation}
\label{eq. A1}
\widehat{H}_{F}=\int dk\hbar \omega (k)\widehat{a}^{\dagger }(k)\widehat{a}(k)-\int d^{3}{\mathbf{x}}\left\{ \left[ \frac{\Delta \chi ^{(1)}({\mathbf{x}})}{2\epsilon (\omega _{0})}\right] :|\widehat{\mathbf{D}}|^{2}({\mathbf{x}}):+\, \left[ \frac{\chi ^{(3)}({\mathbf{x}})}{4\epsilon ^{3}(\omega _{0})}\right] :|\widehat{\mathbf{D}}|^{4}({\mathbf{x}}):\right\} \, .
\label{Ham_f}
\end{equation}
 Here \( \omega (k) \) is the angular frequency of modes with wave-vector \( k \),
describing the \textit{linear} polariton excitations in the fiber, including
dispersion. We will assume that \( \omega (k) \) describes the average linear
response of the fiber, in the limit of a spatially uniform environment. If the
fiber is spatially nonuniform, then it is necessary to add additional inhomogeneous
terms to the Hamiltonian, of generic form \( \Delta \chi ^{(1)}(\mathbf{x}) \).
As usual, \( \epsilon (\omega _{0}) \) is the mode-average dielectric permittivity
at a carrier frequency \( \omega _{0}=\omega (k_{0}), \) while \( \widehat{a}(k) \)
is an annihilation operator defined so that 
\begin{equation}
\label{eq. A2}
\left[ \widehat{a}(k'),\widehat{a}^{\dagger }(k)\right] =\delta (k-k')\, \, .
\end{equation}
 The coefficient \( \chi ^{(3)}(\mathbf{x}) \) is the nonlinear coefficient
arising when the electronic polarization field is expanded as a function of
the electric displacement, in the commonly used Bloembergen\cite{Bloembergen}
notation (the units are S.I. units, following current standard usage). This
may vary along the longitudinal position on the fiber, if the fiber has a variable
composition. In terms of modes of the waveguide, and neglecting modal dispersion,
the electric displacement field operator \( \widehat{\mathbf{D}}(\mathbf{x}) \)
is: 
\begin{eqnarray}
{\widehat{\mathbf{D}}({\mathbf{x}})}=i\int dk\left[ \frac{\hbar k\epsilon (\omega (k))v(k)}{4\pi }\right] ^{\frac{1}{2}}\widehat{a}(k){\mathbf{u}}({\mathbf{r}})\exp(ikx)+h.c.\, \, ,
\end{eqnarray}
 where: 
\begin{equation}
\label{eq. A3}
\int d^{2}{\mathbf{r}}|{\mathbf{u}}({\mathbf{r}})|^{2}=1\, \, .
\end{equation}
 Here \( v(k)=\partial \omega (k)/\partial k \) is the group velocity. The function \( {\mathbf{u}}({\mathbf{r}}) \) gives the transverse mode structure.  Although a general mode structure can be included, for the purposes of this paper we could equally well assume a square wave-guide of area \( A \), which gives
\( {\mathbf{u}}({\mathbf{r}})\simeq {\mathbf{e}}_{y}/\sqrt{A} \). We note here
that the above mode expansion for a dispersive medium is a rather general one,
and has been worked out both from macroscopic quantization\cite{s208}, and
from a microscopic model\cite{Drumhill} with an arbitrary number of electronic
or phonon resonances.

In the infinite volume limit, the polariton field is defined by noting that
the annihilation and creation operators can be related to a quantum field using:
\begin{equation}
\label{eq. A4}
\widehat{\Psi }(t,x)=\frac{1}{\sqrt{2\pi }}\int dk\, \widehat{a}(t,k)\exp[i(k-k_{0})x+i\omega _{0}t]\, \, .
\end{equation}
 This photon-density operator \( \widehat{\Psi }(t,x) \) is the slowly varying
field annihilation operator for the linear quasi-particle excitations of the
coupled electromagnetic and polarization fields traveling through the fiber\cite{s21}.
The nonzero equal-time commutations relations for these Bose operators are
\begin{eqnarray}
\left[ \widehat{\Psi }(t,x),\widehat{\Psi }^{\dagger }(t,x')\right] =\delta (x-x')\, \, .
\end{eqnarray}

As shown in earlier treatments\cite{2}, the Hamiltonian [Eq.~(\ref{Ham_f})] can now be rewritten approximately as: 
\begin{equation}
\widehat{H}_{F}=\hbar \int dx\int dx'\omega (x,x')\widehat{\Psi }^{\dagger }(t,x)\widehat{\Psi }(t,x')-\frac{\hbar }{2}\int dx\chi ^{E}(x)\widehat{\Psi }^{\dagger 2}(t,x)\widehat{\Psi }^{2}(t,x)\, \, .
\label{ham_2f}\end{equation}
Here we have introduced the kernel \( \omega (x,x') \), which is the linear
dielectric component of the Hamiltonian, and a nonlinear coupling term $\chi_E(x)$. This kernel is then Taylor expanded around \( k=k_{0} \), and approximated to quadratic order in \( (k-k_{0}) \), by:
\begin{eqnarray}
\omega (x,x') & = & \int \frac{dk}{2\pi }\omega (k)\exp[i(k-k_{0})(x-x')]-\frac{1}{2}k_{0}v(k_{0})\int d^{2}{\mathbf{r}}\Delta \chi ^{(1)}({\mathbf{x}})|{\mathbf{u}}({\mathbf{r}})|^{2}\delta (x-x')\nonumber \\
 & \simeq  & [\omega _{0}+\Delta \omega (x)]\delta (x-x')+\int \frac{dk}{4\pi }\big [i\omega' _{0}(\partial _{x'}-\partial _{x})+\omega'' _{0}(\partial _{x}\partial _{x'})+\cdots \big ]\exp[ik(x-x')]\, \, .
\end{eqnarray}

In writing down Eq.~(\ref{ham_2f}), we have assumed that the frequency dependence in the nonlinear coupling can be neglected, which is a good approximation for relatively narrow band-widths. The nonlinear term
is often called the \( \chi ^{(3)} \) effect, so named because it arises from
the third order term in the expansion of the polarization field in terms of
the electric field\cite{s05}. It causes an electronic contribution \( n_{2e} \)
to the intensity dependent refractive index, where: \( n=n_{0}+In_{2}=n_{0}+I(n_{2e}+n_{2p}) \).
Thus we define \( \chi ^{E} \), in units of \( [m/s] \), as: 
\begin{equation}
\label{eq. A5}
\chi ^{E}(x)\equiv \left[ \frac{3\hbar w^{2}_{0}v(k_{0})^{2}}{4\epsilon (\omega _{0})c^{2}}\right] \int d^{2}{\mathbf{r}}\chi ^{(3)}({\mathbf{x}})|{\mathbf{u}}({\mathbf{r}})|^{4}\equiv \left[ \frac{\hbar n_{2e}(x)\omega _{0}^{2}v^{2}}{{A}c}\right] \, \, .
\end{equation}
 Here \( {A}=[\int d^{2}{\mathbf{r}}|{\mathbf{u}}({\mathbf{r}})|^{4}]^{-1} \)
is the effective modal cross-section, and \( n_{2e} \) is the refractive index
change per unit field intensity due to electronic transitions. This is less
than the total value observed for \( n_{2} \), since phonon contributions have
yet to be included.

The free evolution part of the total Hamiltonian, which will be removed in subsequent calculations, just describes the carrier frequency \( \omega _{0} \).
This is not needed in Heisenberg picture calculations for \( \widehat{\Psi }(t,x) \),
since it is cancelled by the slowly varying field definition. Next, on partial
integration of the derivative terms and Fourier transforming, the resulting
interaction Hamiltonian \( \widehat{{H}}_{F}' \) describing the evolution of
\( \widehat{\Psi } \) in the slowly varying envelope and rotating-wave approximations
is:
\begin{eqnarray}
\widehat{{H}}_{F}' & = & \widehat{{H}}_{F}-\int dk\hbar \omega _{0}\widehat{a}^{\dagger }(k)\widehat{a}(k)\nonumber \\
 & = & \frac{\hbar }{2}\int _{-\infty }^{\infty }dx\left[ \Delta \omega (x)\widehat{\Psi }^{\dagger }\widehat{\Psi }+\frac{iv}{2}\left( \nabla \widehat{\Psi }^{\dagger }\widehat{\Psi }-\widehat{\Psi }^{\dagger }\nabla \widehat{\Psi }\right) +\frac{\omega ''}{2}\nabla \widehat{\Psi }^{\dagger }\nabla \widehat{\Psi }-\frac{\chi ^{E}(x)}{2}\widehat{\Psi }^{\dagger 2}\widehat{\Psi }^{2}\right] \, \, .
\label{ham_fd}
\end{eqnarray}

For simplicity, only quadratic dispersion is included here. However, the extension
to higher-order dispersion is relatively straightforward. This can be achieved
by including higher-order terms in the expansion of the dielectric kernel, or
else by treating the dispersion as part of the reservoir response function -
as in following sections. The response function approach has the advantage that
a completely arbitrary polarization response can be included, and transformations
to a different frame of reference are simplified. If part of the dielectric
response is treated using response functions, then this part of the measured
refractive index must be excluded from the free Hamiltonian, to avoid double-counting.

\subsection{Heisenberg equation}

From the interaction Hamiltonian [Eq.~\ref{ham_fd}], we find the following Heisenberg
equation of motion for the field operator propagating in the \( +x \) direction:
\begin{equation}
\label{eq. A6}
\left( v\frac{\partial }{\partial x}+\frac{\partial }{\partial t}\right) \widehat{\Psi }(t,x)=\left[ -i\Delta \omega (x)+\frac{i\omega ''}{2}\frac{\partial ^{2}}{\partial x^{2}}+i\chi ^{E}(x)\widehat{\Psi }^{\dagger }(t,x)\widehat{\Psi }(t,x)\right] \widehat{\Psi }(t,x)\quad .
\end{equation}
 This is the quantum nonlinear Schr\"{o}dinger equation in the laboratory frame
of reference, 
which is completely equivalent to the theory of a
 Bose gas of massive quasi-particles with an effective mass of \(\hbar/\omega ''  \) and an average velocity
of \( v \), for photons near to the carrier frequency of interest. It includes
the possibility that the dielectric constant (i.e., the linear response) has
a spatial variation, through the term \( \Delta \omega (x) \) .

We note here that it is occasionally assumed that operators obey equal-space,
rather than equal-time commutation relations. This cannot be exactly true, since
commutators in an interacting quantum field theory are only well-defined at
equal times. At different times, it is possible for a causal effect to propagate
to a different spatial location, which can therefore change the unequal-time
commutator. In other words, imposing free-field equal-space but unequal-time
commutators is not strictly compatible with causality. The assumption of equal-space
commutators may be used as an approximation under some circumstances, provided
interactions are weak. In this paper, we will use standard equal-time commutators.

\section{Raman Hamiltonian}

\label{RH} To the Hamiltonian given in Eq.~(\ref{ham_fd}) must be added couplings to linear
gain, absorption and phonon reservoirs\cite{CD,s55,s76}. The gain and absorption
reservoirs are discussed at length in section \ref{GA}. The phonon field consists
of thermal and spontaneous excitations in the displacement of atoms from their
mean locations in the dielectric lattice. Although previous quantum treatments
of Raman scattering have been given \cite{s34}, it is necessary to modify these
somewhat in the present situation. The Raman interaction energy\cite{Levenson,CD}
of a fiber, in terms of atomic displacements from their mean lattice positions,
is known to have the form: 
\begin{eqnarray}
{H}_{R}=\frac{1}{2}\sum _{j}\eta _{j}^{R}\vdots {\mathbf{D}}(\bar{\mathbf{x}}^{j}){\mathbf{D}}(\bar{\mathbf{x}}^{j})\delta {\mathbf{x}}^{j}+\frac{1}{2}\sum _{ij}\kappa _{ij}:\delta {\mathbf{x}}^{i}\delta {\mathbf{x}}^{j}\, \, .\label{6} 
\end{eqnarray}
 Here \( {\mathbf{D}}(\bar{\mathbf{x}}^{j}) \) is the electric displacement
at the j-th mean atomic location \( \bar{\mathbf{x}}^{j},\; \delta {\mathbf{x}}^{j} \)
is the atomic displacement operator, \( \eta _{j}^{R} \) is a Raman coupling
tensor, and \( \kappa _{ij} \) represents the short-range atom-atom interactions.

In order to quantize this interaction with atomic positions using our macroscopic
quantization method, we must now take into account the existence of a corresponding
set of phonon operators, \( \widehat{b}(\omega ,x) \) and \( \widehat{b}^{\dagger }(\omega ,x) \).
These operators diagonalize the atomic displacement Hamiltonian in each fiber segment,
and have well-defined eigen-frequencies. There are calculations\cite{BellDean70}
of the frequency spectrum and normal modes of vibration for vitreous silica,
using physical models based on the random network theory of disordered systems.
The computed vibrational frequency spectrum is remarkably similar to the observed
Raman gain profile\cite{e78}. The phonon-photon coupling induces Raman transitions
and scattering from acoustic waves (the Brillouin effect) resulting in extra
noise sources and an additional contribution to the nonlinearity. The initial
state of phonons is thermal, with \( n_{th}(\omega )=\left[ \exp {(\hbar \omega /kT)}-1\right] ^{-1} \).

\subsection{Hamiltonian and Heisenberg equations}

In terms of these phonon operators, the fiber Hamiltonian in the interaction
picture and within the rotating wave approximation for a single polarization
is\cite{CD} \( \widehat{{H}}'=\widehat{{H}}_{R}+\widehat{{H}}_{F}' \), where
we have introduced a Raman interaction Hamiltonian: 
\begin{equation}
\widehat{{H}}_{R}=\hbar \int _{-\infty }^{\infty }dx\int _{0}^{\infty }d\omega \left\{ \widehat{\Psi }^{\dagger }(x)\widehat{\Psi }(x)R(\omega ,x)\left[ \widehat{b}(\omega ,x)+\widehat{b}^{\dagger }(\omega ,x)\right] +\omega \widehat{b}^{\dagger }(\omega ,x)\widehat{b}(\omega ,x)\right\} \, \, .
\end{equation}
 Here, the atomic vibrations within the silica structure of the fiber are modeled
as a continuum of localized oscillators, and are coupled to the radiation modes
by a Raman transition with a real frequency dependent coupling \( R(\omega ,x) \).
This coupling could be nonuniform in space, and is determined empirically through
measurements of the Raman gain spectrum\cite{CD}. The atomic displacement is
proportional to \( \widehat{b}+\widehat{b}^{\dagger } \), where the phonon
annihilation and creation operators, \( \widehat{b} \) and \( \widehat{b}^{\dagger } \),
have the equal-time commutations relations 
\begin{eqnarray}
\left[ \widehat{b}(t,\omega ,x),\widehat{b}^{\dagger }(t,\omega ',x')\right] =\delta (x-x')\delta (\omega -\omega' )\, \, .
\end{eqnarray}
 Thus the Raman excitations are treated as an inhomogeneously broadened continuum
of modes, localized at each longitudinal location \( x \). GAWBS (Guided Wave
Acoustic Brillouin Scattering)\cite{s795,s80,s109,s20} is a special case of this, in the
low-frequency limit. Since neither Raman nor Brillouin excitations are completely
localized, this treatment requires a frequency and wave-number cut-off, so that
the field operator \( \widehat{\Psi } \) is slowly varying on the phonon scattering
distance scale.  The corresponding coupled nonlinear operator equations are: 
\begin{eqnarray}
\left( v\frac{\partial }{\partial x}+\frac{\partial }{\partial t}\right) \widehat{\Psi }(t,x) & =i & \left[ -\Delta \omega (x)+\frac{\omega'' }{2}\frac{\partial ^{2}}{\partial x^{2}}+\chi ^{E}(x)\widehat{\Psi }^{\dagger }(t,x)\widehat{\Psi }(t,x)\right] \widehat{\Psi }(t,x)\nonumber \\
 & - & i\left\{ \int ^{\infty }_{0}R(\omega ,x)\left[ \widehat{b}(t,\omega ,x)+\widehat{b}^{\dagger }(t,\omega ,x)\right] d\omega \right\} \widehat{\Psi }(t,x)\, \, ,\nonumber \\
\frac{\partial }{\partial t}\widehat{b}(t,\omega ,x) & = & -i\omega \widehat{b}(t,\omega ,x)-iR(\omega ,x)\widehat{\Psi }^{\dagger }(t,x)\widehat{\Psi }(t,x)\, \, .
\end{eqnarray}

In summary, the original theory of nonlinear quantum field propagation is now
extended to include both the the electronic
and the Raman nonlinearities. The result is a modified Heisenberg equation with a delayed
nonlinear response to the field due to the Raman coupling. On integrating the
Raman reservoirs, one obtains: 
\begin{eqnarray}
\left( v\frac{\partial }{\partial x}+\frac{\partial }{\partial t}\right) \widehat{\Psi }(t,x) & = & i\left[-\Delta \omega (x)+\frac{\omega ^{\prime \prime }}{2}\, \frac{\partial ^{2}}{\partial {x}^{2}}+\int _{0-}^{\infty }dt'\, \chi (t',x)[\widehat{\Psi }^{\dagger }\widehat{\Psi }](t-t',x)+\widehat{\Gamma }^{R}(t,x)\right]\widehat{\Psi }(t,x)\, ,\nonumber 
\end{eqnarray}
where 
\begin{eqnarray}
\chi (t,x) & = & \chi ^{E}(x)\delta (t)+2\Theta (t)\int _{0}^{\infty }R^{2}(\omega ,x)\sin (\omega t)d\omega \nonumber \\
\widehat{\Gamma }^{R}(t,x) & = & -\int ^{\infty }_{0}R(\omega ,x)\left[ \widehat{b}(t,\omega ,x)+\widehat{b}^{\dagger }(t,\omega ,x)\right] d\omega, 
\end{eqnarray}
in which we have defined \( \Theta (t) \) as the step function.

The operators $\widehat{\Gamma }^{R}$ and $\widehat{\Gamma }^{R\dagger}$ are stochastic, with Fourier transforms
defined using the normal Fourier transform conventions for field operators:
\begin{eqnarray}
\widehat{\Gamma }^{R}(\omega ,x) & = & \frac{1}{\sqrt{2\pi }}\int dt\exp (i\omega t)\widehat{\Gamma }^{R}(t,x)\\
\widehat{\Gamma }^{R\dagger }(\omega ,x) & = & \frac{1}{\sqrt{2\pi }}\int dt\exp (-i\omega t)\widehat{\Gamma }^{R}(t,x).
\end{eqnarray}
The frequency-space correlations are given by: 
\begin{eqnarray}
\langle \widehat{\Gamma }^{R\dagger }(\omega ',x')\, \widehat{\Gamma }^{R}(\omega ,x)\rangle  & = & 2\chi ^{\prime \prime }(x,|\omega |)\left[n_{th}({|\omega |})+\Theta (-\omega )\right]{\delta (x-x')}{\delta (\omega -\omega' )}.
\end{eqnarray}
In this expression, we introduce a Raman amplitude gain of \( \chi ^{\prime \prime } \)
per unit photon flux, equal to the imaginary part of the Fourier transform of
\( \chi (t,x) \), so that: \( \chi ^{\prime \prime }(x,|\omega |)=\pi R^{2}(x,|\omega |) \).
Here we use the Bloembergen normalization for response function Fourier transforms,
\begin{equation}
\widetilde{\chi }(\omega ,x)=\int dt\exp (i\omega t)\chi (t,x)\, \, ,
\end{equation}
 which does not have the \( \sqrt{2\pi } \) factor included.
 
 Of some significance is the physical interpretation of the correlation functions,
which can be regarded as directly contributing to the normally ordered spectrum
of the transmitted field. Given a cw carrier, the correlations when \( \omega  \) is positive
correspond to an anti-Stokes (blue-shifted) spectral term, which is clearly
zero unless the thermal phonon population is appreciable. However, when \( \omega  \) is negative, the theta function term indicates that the Stokes (red-shifted) spectral term is nonzero, due to spontaneous Stokes photons emitted even at zero temperature.

\subsection{Raman gain measurements}

The measured intensity gain due to Raman effects at a given relative frequency
\( \omega  \) per unit length, per unit carrier photon flux \( I_{0}=v\langle \widehat{\Psi }^{\dagger }\widehat{\Psi }\rangle  \),
is:
\begin{equation}
\frac{1}{I_{0}}\frac{\partial \ln I}{\partial x}=-2\chi ^{\prime \prime }(\omega ,x)/v^{2}\, .
\end{equation}
Here the gain is positive for Stokes-shifted frequencies (\( \omega <0 \)
), and negative for anti-Stokes (\( \omega >0 \)), as one would expect. This
relationship allows the coupling to be estimated from measured Raman gain and
fluorescence properties. The simplest way to achieve this goal is to expand
the Raman response function in terms of a multiple-Lorentzian model, which can
then be fitted to observed Raman fluorescence data using a nonlinear least-squares
fit. We therefore expand: 
\begin{eqnarray}
\chi (t,x)=\chi ^{E}(x)\delta (t)+\chi (x)\Theta (t)\sum _{j=0}^{n}F_{j}\delta _{j}\exp(-\delta _{j}t)\sin (\omega _{j}t)\, . & 
\end{eqnarray}
For normalization purposes, we have introduced \( \chi (x) \), which is defined
as the total effective nonlinear phase-shift coefficient per unit time and photon
density (in units of \( rad.m/s \)), obtained from the low-frequency nonlinear
refractive index. This is given in terms of the electronic or fast-responding
nonlinear coefficient, \( \chi ^{E}(x) \) , together with the Raman contribution,
by integrating over time: 
\begin{equation}
\label{chie}
\chi (x)=\chi ^{E}(x)+2\int ^{\infty }_{0}\int ^{\infty }_{0}R^{2}(\omega ,x)\sin (\omega t)d\omega dt\, .
\end{equation}

In the above expansion, \( F_{j} \) are a set of dimensionless Lorentzian
strengths, and \( \omega _{j} \) and \( \delta _{j} \) are the resonant frequencies
and widths respectively, of the effective Raman resonances at each frequency.
To improve convergence, the Lorentzian strength parameters are not constrained
to be positive. The \( j=0 \) Lorentzian models the Brillouin contribution
to the response function. In general, all of these parameters could be \( x \)-dependent,
but we will often assume that they are constant in space for notational simplicity.
The values for an \( n=10 \) fit 
in the case of a typical fused silica fiber 
are given in Table \ref{lorfit},
including an estimate of the effective Brillouin contribution averaged
over the individual Brillouin scattering peaks.  
The coefficient
of the electronic nonlinearity is now obtained explicitly in terms of the total
nonlinear refractive index: 
\begin{eqnarray}
\chi ^{E}(x)=\frac{\hbar (1-f)n_{2}\omega _{0}^{2}v^{2}}{{A}c}, & 
\end{eqnarray}
 where \( \omega _{0} \) is the carrier frequency, \( {A} \) is the effective
cross-sectional area of the traveling mode, and \( f \) is the fraction of
the nonlinearity due to the Raman gain, which has been estimated using the procedure
outlined above: 
\begin{eqnarray}
f & = & \frac{\chi ^{R}}{\chi }=\frac{2}{\chi }\int _{0}^{\infty }dt\int _{0}^{\infty }d\omega R^{2}(\omega ,x)\sin (\omega t)\nonumber \\
 & \simeq  & 0.2.
\end{eqnarray}

A result of this model is that the phonon operators do not have white noise
behaviour. In fact, this colored noise property is significant enough to invalidate
the usual Markovian and rotating-wave approximations, which are therefore not
used in the phonon equations. Of course, the photon modes may also be in a thermal
state of some type. However, thermal effects are typically much more important
at the low frequencies that characterize Raman and Brillouin scattering, than
they are at optical frequencies. In addition, if the input is a photon field
generated by a laser, any departures from coherent statistics will be rather
specific to the laser type, instead of having the generic properties of thermal
fields.

Finally, there is another effect which has been so far neglected. This is the ultra-low
frequency tunneling due to lattice defects\cite{Perlmutter}. As this is not strictly
linear, it can not be included accurately in our macroscopic Hamiltonian. Despite
this, the effects of this \( 1/f \) type noise may be included approximately
for any predetermined temperature. This can be achieved by simply modifying
the refractive index perturbation term so that it becomes \( \Delta \omega (t,x) \)
and generates the known low-frequency refractive index fluctuations.

\section{Gain and absorption}

\label{GA} In silica optical fibers, there is a relatively flat absorption
profile, with a minimum absorption coefficient of approximately \( 0.2dB/km \)
in the vicinity of the commonly used communications wavelengths of around \( \lambda =1.5\mu m \).
This effect can be compensated for by the use of fiber laser amplifiers, resulting
in nearly zero net absorption over a total link that includes both normal and
amplified fiber segments. In practical terms, this situation leads to an approximately
uniform fiber environment, provided the net gain and loss are spatially modulated
more rapidly than than the pulse dispersion length. These additional effects
need to be included within the present Hamiltonian model, in order to have a
fully consistent quantum theory. For wide-band communications systems, either
with time-domain multiplexing or frequency-domain multiplexing, it can become
necessary to include the frequency-dependence and spatial variation of the gain
and loss terms. This is especially true if spectral filters are included in
the fiber.

\subsection{Absorbing reservoirs}

The absorption reservoir is modeled most simply by a coupling to a continuum
of harmonic oscillators at resonant frequency \( \omega . \) In the interaction
picture used here, the Hamiltonian term causing rapidly varying operator evolution
of the reservoir at the carrier frequency \( \omega _{0} \) is subtracted,
leaving: 
\begin{equation}
\widehat{H}_{A}'=\hbar \int _{-\infty }^{\infty }dx\int _{0}^{\infty }d\omega \left\{ [\widehat{\Psi }(x)\widehat{a}^{\dagger }(\omega ,x)A(\omega ,x)+h.c.\, ]+(\omega -\omega _{0})\widehat{a}^{\dagger }\widehat{a}(\omega ,x)\right\} \, \, ,
\end{equation}
 where $A(\omega, x)$ provides the frequency dependent coupling between the radiation modes and the absorption reservoirs.  The reservoir annihilation and creation operators, \( \widehat{a} \)
and \( \widehat{a}^{\dagger } \), have the commutation relations 
\begin{eqnarray}
\left[ \widehat{a}(\omega ,x),\widehat{a}^{\dagger }(\omega ',x')\right] =\delta (x-x')\delta (\omega -\omega' )\, \, .
\end{eqnarray}

The equations for the absorbing photon reservoirs can be integrated immediately.
The photon reservoir variable, for instance, obeys: 
\begin{equation}
\frac{\partial }{\partial t}\widehat{a}(t,\omega ,x)=-i(\omega -\omega _{0})\widehat{a}(t,\omega ,x)-iA(\omega ,x)\widehat{\Psi }(t,x)\, \, .
\end{equation}
 Hence, the solutions are: 
\begin{equation}
\widehat{a}(t,\omega ,x)=\widehat{a}(t_{0},\omega ,x)\, \exp[-i(\omega -\omega _{0})(t-t_{0})]-iA(\omega ,x)\int ^{t}_{t_{0}}\exp[-i(\omega -\omega _{0})(t-t')]\widehat{\Psi }(t',x)dt'\, \, ,
\end{equation}
 with initial correlations for the reservoir variables in the far past \( (t_{0}\, \rightarrow \, -\infty ) \)
given by: 
\begin{eqnarray}
\langle \widehat{a}^{\dagger }(t_{0},\omega ,x)\widehat{a}(t_{0},\omega ',x')\rangle  & = & n_{th}(\omega )\delta (x-x')\delta (\omega -\omega' )\, \, ,\nonumber \\
\langle \widehat{a}(t_{0},\omega ,x)\widehat{a}^{\dagger }(t_{0},\omega ',x')\rangle  & = & [n_{th}(\omega )+1]\delta (x-x')\delta (\omega -\omega' )\, \, .
\end{eqnarray}

The solution for \( \widehat{a}(t,\omega ,x) \) is substituted into the Heisenberg
equation for the field evolution, giving rise to an extra time-dependent term,
of the form: 
\begin{eqnarray}
-i\int _{0}^{\infty }A^{*}(\omega ,x)\widehat{a}(t,\omega ,x)d\omega  & = & -\int _{0}^{\infty }d\omega |A(\omega ,x)|^{2}\int ^{t}_{t_{0}}dt'\exp[-i(\omega -\omega _{0})(t-t')]\widehat{\Psi }(t',x)\nonumber \\
 & - & i\int _{0}^{\infty }d\omega A^{*}(\omega ,x)\exp[-i(\omega -\omega _{0})(t-t_{0})]\widehat{a}(t_{0},\omega ,x)\nonumber \\
 & = & -\int _{0}^{\infty }dt''\, \gamma ^{A}(t'',x)\widehat{\Psi }(t-t'',x)+\widehat{\Gamma }^{A}(t,x)\, ,
\end{eqnarray}
 where \( t''=t-t' \) , and the response function and reservoir terms are obtained
most simply by extending the lower limit on the frequency integral to \( -\infty , \)
introducing only an infinitesimal error in the process, so that: 
\begin{eqnarray}
\gamma ^{A}(t,x) & \approx  & \Theta (t)\int _{-\infty }^{+\infty }d\omega \, |A(\omega ,x)|^{2}\, \exp[-i(\omega -\omega _{0})t]\nonumber \\
\widehat{\Gamma }^{A}(t,x) & = & -i\int _{0}^{\infty }d\omega A^{\dagger }(\omega ,x)\exp[-i(\omega -\omega _{0})(t-t_{0})]\widehat{a}(t_{0},\omega ,x).\label{ga_res}
\end{eqnarray}

The response function integral represents a deterministic or `drift' term to
the motion, with a Fourier transform of: 
\begin{equation}
\widetilde{\gamma }^{A}(\omega ,x)=\int \gamma ^{A}(t,x)\exp(i\omega t)dt=\gamma ^{A}(\omega ,x)+i\gamma ^{A\prime \prime }(\omega ,x)\, \, ,
\end{equation}
 so that the amplitude loss rate is:
\begin{equation}
\gamma ^{A}(\omega ,x)={\pi }\, |A(\omega _{0}+\omega ,x)|^{2}\, \, .
\end{equation}
In the case of a spatially uniform reservoir with a flat spectral density, the
Wigner-Weiskopff approximation (neglecting frequency shifts) gives a uniform
Markovian loss term with: 
\begin{eqnarray}
\gamma ^{A}(t) & \approx  & \widetilde{\gamma }^{A}\delta (t)\, \, ,
\end{eqnarray}
 where the average amplitude loss coefficient is: 
\begin{eqnarray}
\widetilde{\gamma }^{A}=\widetilde{\gamma }^{A}(0) & = & \int _{0}^{\infty }\int _{-\infty }^{+\infty }dtd\omega \, |A(\omega )|^{2}\, \exp[-i(\omega -\omega _{0})t]\nonumber \\
 & = & \gamma ^{A}+i\gamma ^{A\prime \prime }\, .
\end{eqnarray}
This approximation, known as the Markov approximation, is generally rather accurate
for the absorbing reservoirs, whose response does not typically vary fast with
frequency. An exception to this rule would be any case involving resonant impurities
in the fiber, or very short pulses whose bandwidth is comparable to the frequency-scale
of absorption changes.

The second quantity in Eq.~(\ref{ga_res}), \( \widehat{\Gamma }^{A}(t,x) \), behaves like a stochastic
term due to the random initial conditions. Neglecting the frequency dependence
of the thermal photon number, the corresponding correlation functions are 
\begin{eqnarray}
{\langle }\widehat{\Gamma }^{A}(t,x)\widehat{\Gamma }^{A\dagger }(t',x'){\rangle }\;  & = & \int _{0}^{\infty }d\omega |A(\omega ,x)|^{2}\exp[-i(\omega -\omega _{0})(t-t')][n_{th}(\omega )+1]\delta (x-x')\nonumber \\
 & \approx  & [\gamma ^{A}(t-t',x)+\gamma ^{A*}(t'-t,x)][n_{th}(\omega _{0})+1]\, \delta (x-x')\, ,
\end{eqnarray}
 and: 
\begin{eqnarray}
{\langle }\widehat{\Gamma }^{A\dagger }(t',x')\widehat{\Gamma }^{A}(t,x){\rangle }\;  & = & \int _{0}^{\infty }d\omega |A(\omega ,x)|^{2}\exp[-i(\omega -\omega _{0})(t-t')]n_{th}(\omega )\delta (x-x')\nonumber \\
 & \approx  & [\gamma ^{A}(t-t',x)+\gamma ^{A*}(t'-t,x)]n_{th}(\omega _{0})\, \delta (x-x')\, .
\end{eqnarray}
At optical or infra-red frequencies, it is a good approximation to set \( n_{th}(\omega _{0})=0 \).
On Fourier transforming the noise sources, one then obtains: 
\begin{equation}
{\langle }\widehat{\Gamma }^{A}(\omega ,x)\widehat{\Gamma }^{A\dagger }(\omega ',x'){\rangle }=2\gamma ^{A}(\omega ,x)\delta ({x}-{x}^{\prime })\delta ({\omega }-{\omega }^{\prime }).
\end{equation}
Again taking the simplifying case of a spatially uniform reservoir in the Wigner-Weiskopff
limit, this reduces to: 
\begin{eqnarray}
{\langle }\widehat{\Gamma }^{A}(t,x)\widehat{\Gamma }^{A\dagger }(t',x'){\rangle }\;  & = & 2\gamma ^{A}\delta (t-t')\delta (x-x')\nonumber \\
{\langle }\widehat{\Gamma }^{A\dagger }(t,x)\widehat{\Gamma }^{A}(t',x'){\rangle }\;  & = & 0.
\end{eqnarray}

Note that the dimensions for the amplitude relaxation rates are \( [\gamma ^{A}]\, =\, s^{-1}. \)
It is easy to show that \( 2\gamma ^{A}/v \) corresponds to the usual linear
absorption coefficient for fibers during propagation. A typical measured absorption
figure in current fused silica communications fibers is \( 0.2dB/km \) in the
minimum region of absorption (near \( \lambda =1.5\mu m \)). The corresponding
absorption coefficient is \( 2\gamma ^{A}/v\, \simeq \, 2.3\times 10^{-5}m^{-1}. \)
To the extent that this effect is wavelength (and hence frequency) dependent,
the resulting dispersion can be included as well, giving rise to a complete
response function \( \gamma ^{A}(t) \) for absorption. Non-Markovian effects
like this can either be neglected completely -- which is a good approximation for slowly varying
absorption in undoped fiber -- or else included in the correlation functions of the
reservoirs as given above.

The physical meaning of the reservoir operator spectral correlations is best
understood by considering the effect of these terms on photodetection,
which according to photodetection theory, means a normally ordered field
correlation. This involves normally ordered reservoir correlations to lowest
order. Since these are zero, we conclude that the absorbing reservoirs essentially
add no quantum noise that is observable via normal photodetection methods.

\subsection{Waveguide laser amplifiers}

\label{sec:Amp} The equations for gain or laser reservoirs are generally more
complex, involving the nonlinear response of atomic impurities added to provide
some gain in the fiber medium. This also involves a pump process (usually from
a semiconductor laser) to maintain the lasing atoms in an inverted state. In
the case of silica fibers, a commonly used lasing transition is provided by 
erbium impurities\cite{Mears}. The effect of these gain reservoirs
is typically to introduce new types of dispersion, owing to the frequency dependence
of the gain\cite{Desurvire}. In addition, there are new nonlinear effects,
due to the effects of saturation - which in turn depend on the pumping intensity.

It is possible to develop a detailed theory of erbium laser amplifiers. However,
this paper will treat the simplest possible quantum theory of a traveling-wave
quantum-limited laser amplifier. More details of the quantum theory, including
nonlinear effects, are treated elsewhere\cite{DrumRayn}. However, the simple
theory presented here provides a microscopic justification for observed quantum
noise effects in fiber amplifier chains. In particular, it reproduces the results
of the phenomenological theory of Gordon and Haus\cite{s35}, which is known
to give predictions in accord with soliton transmission experiments. The resulting
``Gordon-Haus jitter'' can be reduced through the use of filtering techniques.
Assuming that the laser amplifier is polarization-insensitive, we again omit
the polarization index. The reservoir variable \( \widehat{\sigma }_{\mu }=|1\rangle _{\mu }\langle 2|_{\mu } \)
is an atomic transition operator, which induces a near-resonant atomic transition
from an upper to a lower state, with two-level transitions having an assumed
density of \( \rho (\omega ,x) \) in position and resonant angular frequency
\( \omega  \) .

These quantum effects are modeled here by including gain reservoirs in the Hamiltonian, coupled by a frequency dependent term $G(\omega,x)$ to the radiating field.  Here the
gain terms \( \widehat{\sigma }^{\pm }(\omega ,x,t) \) represent the raising
and lowering Pauli field operators, for two-level lasing transitions at frequency
\( \omega  \). In more detail, we have gain given by an interaction Hamiltonian:
\begin{equation}
\widehat{H}_{G}'=\hbar \int _{-\infty }^{\infty }dx\int _{0}^{\infty }d\omega \left\{ [\widehat{\Psi }\widehat{\sigma }^{+}(\omega ,x)G(\omega ,x)+h.c.]+\frac{\omega -\omega _{0}}{2}\sigma ^{z}(\omega ,x)\right\} \, \, ,
\end{equation}
 where the atomic raising and lowering field operators, \( \widehat{\sigma }^{\pm } \),
are defined in terms of discrete Pauli operators, by: 
\begin{eqnarray}
\widehat{\sigma }^{+}(\omega ,x,t) & = & \frac{1}{\sqrt{\rho (\omega ,x)}}\sum _{\mu }|2\rangle \langle 1|_{\mu }\exp(-i\omega _{0}t)\delta (x-x_{\mu })\delta (\omega -\omega _{\mu })\, ,\nonumber \\
\widehat{\sigma }^{-}(\omega ,x,t) & = & \frac{1}{\sqrt{\rho (\omega ,x)}}\sum _{\mu }|1\rangle \langle 2|_{\mu }\exp(i\omega _{0}t)\delta (x-x_{\mu })\delta (\omega -\omega _{\mu })\, ,\nonumber \\
\widehat{\sigma }^{z}(\omega ,x,t) & = & \frac{1}{\rho (\omega ,x)}\sum _{\mu }[|2\rangle \langle 2|-|1\rangle \langle 1|]_{\mu }\delta (x-x_{\mu })\delta (\omega -\omega _{\mu })\, .
\end{eqnarray}
These operators are in general time-dependent in the Heisenberg picture, but have the equal-time commutation relations:
\begin{eqnarray}
\left[ \widehat{\sigma }^{+}(t,\omega ,x),\widehat{\sigma }^{-}(t,\omega ',x')\right] =\widehat{\sigma }^{z}(t,\omega ,x) & \delta (x-x')\delta (\omega -\omega' ). & 
\end{eqnarray}
 In the limit of complete inversion, with linear response and pure inhomogeneous
broadening, 
\begin{equation}
\frac{\partial }{\partial t}\widehat{\sigma }^{-}(t,\omega ,x)=-i(\omega -\omega _{0})\widehat{\sigma }^{-}(t,\omega ,x)+i\widehat{\sigma }^{z}(t,\omega ,x)G(\omega ,x)\widehat{\Psi }(t,x)\, \, .
\end{equation}
 Hence, the solutions in the amplifier case are: 
\begin{eqnarray}
\widehat{\sigma }^{-}(t,\omega ,x)&=&\widehat{\sigma }^{-}(t_{0},\omega ,x)\, \exp[-i(\omega -\omega _{0})(t-t_{0})]\nonumber \\ &+&iG(\omega ,x)\int ^{t}_{t_{0}}\exp[-i(\omega -\omega _{0})(t-t')]\widehat{\sigma }^{z}(t'\omega ,x)\widehat{\Psi }(t',x)dt'\, \, .
\end{eqnarray}
 With complete inversion, \( \langle \widehat{\sigma }^{z}(t_{0},\omega ,x)\rangle =1 \)
, so the initial correlations for the reservoir variables in the far past \( (t_{0}\, \rightarrow \, -\infty ) \)
are given by: 
\begin{eqnarray}
\langle \widehat{\sigma }^{+}(t_{0},\omega ,x)\widehat{\sigma }^{-}(t_{0},\omega ',x')\rangle  & = & \delta (x-x')\delta (\omega -\omega' )\, \, ,\nonumber \\
\langle \widehat{\sigma }^{-}(t_{0},\omega ,x)\widehat{\sigma }^{+}(t_{0},\omega ',x')\rangle  & = & 0\, \, .
\end{eqnarray}
 
 We substitute the solution for \( \widehat{\sigma }^{-}(t,\omega ,x) \) into
the Heisenberg equation for the field evolution, assuming no depletion of the
inversion, and trace over the atomic gain reservoirs. This gives rise to an
extra time-dependent term, of the form: 
\begin{eqnarray}
-i\int _{0}^{\infty }G^{*}(\omega ,x)\widehat{\sigma }^{-}(t,\omega ,x)d\omega  & = & \int _{0}^{\infty }d\omega |G(\omega ,x)|^{2}\int ^{t}_{t_{0}}dt'\exp[-i(\omega -\omega _{0})(t-t')]\widehat{\Psi }(t',x)\nonumber \\
 & - & i\int _{0}^{\infty }d\omega G^{*}(\omega ,x)\exp[-i(\omega -\omega _{0})(t-t_{0})]\sigma ^{-}(t_{0},\omega ,x)\nonumber \\
 & = & \int _{0}^{\infty }dt''\, \gamma ^{G}(t'',x)\widehat{\Psi }(t-t'',x)+\widehat{\Gamma }^{G}(t,x),
\end{eqnarray}
 where \( t''=t-t' \), as before. This gives: 
\begin{eqnarray}
\gamma ^{G}(t,x) & \approx  & \Theta (t)\int _{-\infty }^{+\infty }d\omega \, |G(\omega ,x)|^{2}\, \exp[-i(\omega -\omega _{0})t] \nonumber \\
\widehat{\Gamma }^{G}(t,x) & \approx  & -i\int _{-\infty }^{\infty }d\omega G^{*}(\omega ,x)\exp[-i(\omega -\omega _{0})(t-t_{0})]\sigma ^{-}(t_{0},\omega ,x)\,\, .
\end{eqnarray}
 Fourier transforming the response function gives: 
\begin{eqnarray}
\widetilde{\gamma }^{G}(\omega ,x)=\int \gamma ^{G}(t,x)\exp(i\omega t)dt=\gamma ^{G}(\omega ,x)+i\gamma ^{G\prime \prime }(\omega ,x)\, \, , & 
\end{eqnarray}
 and the (real) resonant amplitude gain coefficient is: 
\begin{eqnarray}
\gamma ^{G}(\omega ,x)={\pi }\, |G(\omega +\omega _{0},x)|^{2}\, \, .
\end{eqnarray}

As with the loss case, \( \widehat{\Gamma }^{G}(t,x) \) behaves like a stochastic
term due to the random initial conditions. The corresponding correlation functions
are 
\begin{eqnarray}
{\langle }\widehat{\Gamma }^{G\dagger }(t',x')\widehat{\Gamma }^{G}(t,x){\rangle }\;  & = & \int _{0}^{\infty }d\omega |G(\omega ,x)|^{2}\exp[i(\omega -\omega _{0})(t-t')]\delta (x-x')\nonumber \\
 & = & [\gamma ^{G}(t-t',x)+\gamma ^{G^{*}}(t'-t,x)]\, \delta (x-x')\, \, .
\end{eqnarray}
Fourier transforming these noise sources gives: 
\begin{equation}
{\langle }\widehat{\Gamma }^{G\dagger }(\omega ',x')\widehat{\Gamma }^{G}(\omega ,x){\rangle }=2\gamma ^{G}(\omega ,x)\delta ({x}-{x}^{\prime })\delta ({\omega }-{\omega }^{\prime })
\end{equation}
Taking the uniform fiber in the Wigner-Weiskopff limit as before, so \( \gamma ^{G}=\gamma ^{G}(0,x) \),
this reduces to: 
\begin{eqnarray}
{\langle }\widehat{\Gamma }^{G}(t,x)\widehat{\Gamma }^{G\dagger }(t',x'){\rangle }\;  & = & 0\nonumber \\
{\langle }\widehat{\Gamma }^{G\dagger }(t,x)\widehat{\Gamma }^{G}(t',x'){\rangle }\;  & = & 2\gamma ^{G}\, \delta (t-t')\delta (x-x').
\end{eqnarray}

The dimensions for the amplitude gain are \( [\gamma ^{G}]\, =\, s^{-1}. \)
On Fourier transforming, the response function can be related to the measured
intensity gain \( 2{\mbox {\textrm{Re}}}[\widetilde{\gamma }^{G}(\omega ,x)/v] \)
at any frequency offset \( \omega  \), relative to the carrier frequency \( \omega _{0} \).
This allows one to obtain the linear gain coefficient for fibers during propagation.
Since measured laser gain figures can be much greater than the absorption, it
is possible to compensate for fiber absorption with relatively short regions
of gain. The results presented here are only valid in the linear gain regime.
More generally, a functional Taylor expansion up to at least third order in
the field would be needed to represent the full nonlinear response of the laser
amplifier, together with additional quantum noise terms.

Finally, it is necessary to consider the result of incomplete inversion of an
amplifier. Here, the noninverted atoms give rise to absorption, not gain, and
will generate additional quantum-noise and absorption response terms. These
must be treated as in the previous section, including non-Markovian effects
if the absorption line is narrow-band. An important consequence is that the
measured gain only gives the difference between the gain and the loss. This
doesn't cause any problems with the deterministic response function -- but it
does cause difficulties in determining the amplifier quantum noise levels, which
can only be uniquely determined through spontaneous fluorescence measurements.
Obviously, the lowest quantum noise levels occur when all the lasing transitions
are completely inverted.

The physical meaning of the reservoir operator spectral correlations 
for the amplifier case is clearly quite different to the case of the absorber.
If we consider the effect of these terms on photodetection as before,
which means a normally ordered field correlation, we should look again at
the normally ordered correlations of the reservoirs. Since these are 
no longer zero, we conclude that the amplifying reservoirs emit
fluorescent photons due to spontaneous emission over the  amplifier bandwidth.

\section{Combined Heisenberg Equations}

\label{CHE} Coupling linear gain and absorption reservoirs in this way to the 
Raman-modified Heisenberg equation  leads
to a generalized quantum nonlinear Schr\"{o}dinger equation. Such equations are 
sometimes called quantum Langevin
equations. In the present case of a single polarization, the resulting field
equations are:
\begin{eqnarray}
\left( v\frac{\partial }{\partial x}+\frac{\partial }{\partial t}\right) \widehat{\Psi }(t,x) & = & -\int _{0}^{\infty }dt'\, \gamma (t',x)\widehat{\Psi }(t-t',x)+\widehat{\Gamma }(t,x)\nonumber \\
 & + & i\left[\frac{\omega ^{\prime \prime }}{2}\, \frac{\partial ^{2}}{\partial {x}^{2}}+\int _{0-}^{\infty }dt'\, \chi (t')[\widehat{\Psi }^{\dagger }\widehat{\Psi }](t-t',x)+\widehat{\Gamma }^{R}(t,x)\right]\widehat{\Psi }(t,x).
\label{RMQ}\end{eqnarray}
 In this equation, 
\begin{eqnarray}
\gamma (t,x)=\gamma ^{A}(t,x)-\gamma ^{G}(t,x)+i\Delta \omega (x)\delta (t)
\end{eqnarray}
 is a net linear response function due to a coupling to linear gain/absorption
reservoirs, including the effects of a spatially varying refractive index. This
can be Fourier transformed, giving: \( \widetilde{\gamma }(\omega ,x)={\gamma }(\omega ,x)+i\gamma ^{\prime \prime }(\omega ,x) \),
where \( {\gamma }(\omega ,x)<0 \) for gain, and \( {\gamma }(\omega ,x)>0 \)
for absorption. Similarly, \( \widehat{\Gamma }(t,x) \) is the linear quantum
noise due to gain and absorption. The actual measured intensity gain at frequency
\( \omega +\omega _{0} \) is given in units of \( [m^{-1}] \) , by: 
\begin{equation}
\frac{\partial \ln I}{\partial x}=2(\gamma ^{G}(\omega ,x)-\gamma ^{A}(\omega ,x))/v\, .
\end{equation}

The stochastic terms have the correlations
\begin{eqnarray}
\langle \widehat{\Gamma }^{R\dagger }(\omega ',x')\, \widehat{\Gamma }^{R}(\omega ,x)\rangle  & = & 2\chi ^{\prime \prime }(x,|\omega |)[n_{th}({|\omega |})+\Theta (-\omega )]{\delta (x-x')}{\delta (\omega -\omega' )}\nonumber \\
\langle \widehat{\Gamma }^{\dagger }(\omega ',x')\, \widehat{\Gamma }(\omega ,x)\rangle  & = & 2\gamma ^{G}(\omega ,x)\delta ({x}-{x}^{\prime })\delta ({\omega }-{\omega }^{\prime })\nonumber \\
\langle \widehat{\Gamma }(\omega ,x)\, \widehat{\Gamma }^{\dagger }(\omega' ,x')\rangle  & = & 2\gamma ^{A}(\omega ,x)\delta ({x}-{x}^{\prime })\delta ({\omega }-{\omega }^{\prime })\, ,
\end{eqnarray}
where we have introduced minimal linear quantum noise terms $\Gamma$ and $\Gamma^{\dagger}$ for the gain/absorption
reservoirs, and where thermal photons have been neglected (since usually \( \hbar \omega _{0}>>kT \), as explained in section \ref{GA}).  Equation (\ref{RMQ}) can be easily generalized to include nonlinear absorption
or laser saturation effects, relevant to amplifiers with intense fields, but
these terms are omitted here for simplicity.

This complete Heisenberg equation gives a consistent quantum theoretical description
of dispersion, nonlinear refractive index, Raman/GAWBS scattering, linear gain,
and absorption. It is important to notice that the reservoir correlations have
a simple physical interpretation, especially in the zero-temperature limit.
Normally ordered noise correlations occur when there is gain, anti-normally
ordered correlations when there is absorption. This is the reason why the normally ordered
Raman noise correlations vanish at zero temperature for positive frequencies.
At low temperatures, Raman processes only cause absorption to occur at positive
detunings from a pump frequency. Thermal correlations have a more classical
behaviour, and occur for both types of operator ordering.

It is often useful to do calculations in a simpler model, in which we include
the effects of uniform gain and loss in a moving frame. This can either be carried
out using a standard moving frame (\( x_{v}=x-vt \)), or with a propagative
time (\( t_{v}=t-x/v \) ) as in the original Gordon-Haus calculations. For
propagative calculations, it is most convenient to use photon flux operators
\begin{equation}
\widehat{\Phi }(t_{v},x)=\sqrt{v}\widehat{\Psi }(t,x)\, .
\end{equation}
 For long pulses, assuming a uniform gain/loss response in the frequency domain,
the propagative transformation gives the following approximate equations: 
\begin{eqnarray}
\frac{\partial }{\partial x}\widehat{\Phi }(t_{v},x) & = & -\int _{0}^{\infty }dt_{v}'\, \frac{\gamma (t_{v}',x)}{v}\widehat{\Phi }(t_{v}-t_{v}',x)+\widehat{\Gamma }(t)/\sqrt{v}\nonumber \\
 & + & i\biggl [-\frac{k^{\prime \prime }}{2}\, \frac{\partial ^{2}}{\partial {t_{v}}^{2}}+\int _{0-}^{\infty }dt'\, \frac{\chi (t_{v}')}{v^{2}}[\widehat{\Phi }^{\dagger }\widehat{\Phi }](t_{v}-t_{v}',x)+\frac{1}{v}\widehat{\Gamma }^{R}\biggr ]\widehat{\Phi }(t_{v},x)\, .
\end{eqnarray}
In addition, if the pulses are narrow-band compared to the gain and loss bandwidths,
and the reservoirs are uniform, then the gain and absorption reservoirs are
nearly delta-correlated, with 
\begin{eqnarray}
\langle \widehat{\Gamma }^{\dagger }(t,x_{v})\, \widehat{\Gamma }(t',x_{v}')\rangle  & = & 2\gamma ^{G}\delta ({x_{v}}-{x_{v}}^{\prime })\delta (t-t')\nonumber \\
\langle \widehat{\Gamma }(t,x_{v})\, \widehat{\Gamma }^{\dagger }(t',x_{v}')\rangle  & = & 2\gamma ^{A}\delta ({x_{v}}-{x_{v}}^{\prime })\delta (t-t').
\end{eqnarray}

It is essentially this set of approximate equations that corresponds to those
used to predict the soliton\cite{e36} self-frequency shift\cite{s348} and related
effects\cite{s35} in soliton propagation, except for the omission of the Raman reservoir
terms.

\section{Phase-space methods}

\label{PSM} The Heisenberg equations are not readily soluble in their present form.
To generate numerical equations for analytic calculations or for simulation,
operator representation theory can be used. There is more than one possible
method, depending on which phase-space representation is used. The positive-\( P \)
representation, for example, produces exact results\cite{s14,s21,CD} provided phase-space boundary terms are negligible, while a
truncated Wigner representation\cite{s23,s12} gives approximate results 
that are valid in the limit of large photon number.
It is important to note that the Wigner method represents symmetrically ordered
rather than normally ordered operator products, and so has finite quantum noise
terms even for a vacuum field. These can be thought of as corresponding to the
shot noise detected in a homodyne or local-oscillator measurement, while the
positive-\( P \) representation represents normally ordered operators, and
therefore corresponds to direct-detection noise.

Either technique can be used for this problem, each with its characteristic
advantages and disadvantages. The positive-\( P \) representation, although
exact, uses an enlarged phase-space which therefore takes longer to simulate
numerically. It only includes normally ordered noise and initial conditions,
and this is an advantage in some cases, since the resulting noise is zero in
the vacuum state. The Wigner technique is simpler, and for large mode occupations,
its results are accurate enough for many purposes. However, it has the drawback
that it includes symmetrically ordered vacuum fluctuations.

First, we expand the field operators in terms of operators for the free-field
modes. Applying the appropriate operator correspondences to the master equation
for the reduced density operator \( \widehat{\rho }_{\Psi} \) in which the reservoir
modes have been traced over, namely 
\begin{eqnarray}
\widehat{\dot{\rho }}_{\Psi }={\textrm{Tr}}_{R}\widehat{\dot{\rho }}={\textrm{Tr}}_{R}\frac{1}{i\hbar }\left[ \widehat{H},\widehat{\rho }\right] ,
\end{eqnarray}
 gives a functional equation for the corresponding operator representation.

In the positive-$P$ case, the equation is defined on a functional phase-space
of double the classical dimensions, so that a complete expansion in terms of
a coherent-state basis \( |\Psi \rangle  \) is obtained:
\begin{equation}
\hat{\rho }_{\Psi}(t)=\int \int P(t,\Psi ,\overline{\Psi })\, \frac{|\Psi \rangle \langle \overline{\Psi }|}{\langle \overline{\Psi }|\Psi \rangle }d[\Psi ]\, d[\overline{\Psi }]\, .
\end{equation}
 The resulting Fokker-Planck equation for the positive distribution \( P(t,\Psi ,\overline{\Psi }) \)
has only second order derivative terms. Sometimes the notation \( \Psi ^{+}=\overline{\Psi }^{*} \)is
used to indicate the stochastic field that corresponds to the hermitian conjugate
of \( \Psi  \).

The equation for the Wigner function \( W(t,\Psi ) \) also contains third and
fourth order derivative terms, which may be neglected at large photon number.
The resultant Fokker-Planck equation in either case, can be converted into equivalent 
Ito stochastic equations for the phase space variables $\Psi$ (and $\overline{\Psi}$).  
Physical quantities can be calculated by forming the average over many stochastic realizations, or
paths, in phase-space.  For example, in the positive-$P$ representation, $\langle\overline{\Psi}^*\Psi\rangle_{\rm stochastic} =  \langle\widehat\Psi^{\dagger} \widehat\Psi\rangle_{\rm quantum}$, while in the Wigner representation, $\langle\Psi^*\Psi\rangle_{\rm stochastic} =  \frac{1}{2}\langle\widehat\Psi^{\dagger} \widehat\Psi + \widehat\Psi \widehat\Psi^{\dagger}\rangle_{\rm quantum}$.

It should be clear from this that the positive-$P$ representation directly generates an intensity corresponding to the usual normally ordered intensity that is detected in direct photodetection. The Wigner representation, however, generates an intensity result that includes some vacuum fluctuations. In a computer simulation with a finite number $M$ of modes, we must correct the Wigner result by subtracting $M/2$ from any simulated photon number, or $vM/2$ from any calculated photon flux, in order to obtain the direct photodetection result.  For the calculation of a homodyne measurement, the Wigner method will give the most directly suitable result with symmetric ordering. In this case it is the positive-$P$ representation that will need correction terms added to it.  Once these corrections are made, either method will give similar results, although the sampling error may not be identical.

\subsection{Modified nonlinear Schr\"{o}dinger equation}

Standard custom in fiber optics applications\cite{e36} involves using the propagative
reference frame with the normalized variables: \( \tau =(t-x/v)/t_{0} \) and
\( \zeta =x/x_{0} \), where \( t_{0} \) is a typical pulse duration used for
scaling purposes, and \( x_{0}=t_{0}^{2}/|k''|\sim 1km \) for dispersion shifted
fiber. This change of variables is useful only when slowly varying second order
derivatives involving \( \zeta  \) can be neglected, which occurs for \( vt_{0}/x_{0}\ll 1 \).
For typical values of the parameters used in experiments, this inequality is
often well-satisfied (\( vt_{0}\sim 10^{-4}m \)). To make it simpler to compare
with this usage, we will make the same transformation for the stochastic equations
that are equivalent to our complete operator equations, and scale the variables
used in a dimensionless form.

For definiteness, we will now focus on the spatially uniform case. The resultant
equation, which includes gain and loss, is a Raman-modified nonlinear Schr\"{o}dinger
(NLS) equation with stochastic noise terms: 
\begin{eqnarray}
\frac{\partial }{\partial \zeta }\phi (\tau ,\zeta ) & = & -\int ^{\infty }_{-\infty }d\tau' g(\tau -\tau' )\phi (\tau' ,\zeta )+\Gamma (\tau ,\zeta )\nonumber  \\
 & + & \left[\pm \frac{i}{2}\frac{\partial ^{2}\phi }{\partial \tau ^{2}}+i\int ^{\infty }_{-\infty }d\tau' h(\tau -\tau' )\phi ^{*}(\tau' ,\zeta )\phi (\tau' ,\zeta )+\Gamma ^{R}(\tau ,\zeta )\right]\phi (\tau ,\zeta ),
\end{eqnarray}
 where \( \phi =\Psi \sqrt{vt_{0}/\overline{n}} \) is a dimensionless photon
field amplitude. The photon flux is \( |\phi |^{2}\overline{n}/t_{0} \), and
\( \overline{n}=|k''|{A}c/(n_{2}\hbar \omega _{c}^{2}t_{0})=v^{2}t_{0}/\chi x_{0} \)
is the typical number of photons in a soliton pulse of width \( t_{0} \),
for scaling purposes. The positive sign in front of the second derivative term
applies for anomalous dispersion (\( k''<0 \)), which occurs for longer wavelengths,
and the negative sign applies for normal dispersion (\( k''>0 \)). A similar
equation is obtained in the positive-$P$ case, except that \( \phi ^{*} \)and
\( \Gamma ^{R*}(\tau ,\zeta ) \) are replaced by non-complex-conjugate fields
denoted \( \phi ^{+} \) and \( \Gamma ^{R+}(\tau ,\zeta ) \) respectively:
\begin{eqnarray}
\frac{\partial }{\partial \zeta }\phi ^{+} (\tau ,\zeta ) & = & -\int ^{\infty }_{-\infty }d\tau' g^*(\tau -\tau' )\phi ^{+} (\tau' ,\zeta )+\Gamma ^{+} (\tau ,\zeta )\nonumber  \\
 & + & \left[\mp \frac{i}{2}\frac{\partial ^{2}\phi ^{+} }{\partial \tau ^{2}}-i\int ^{\infty }_{-\infty }d\tau' h ^{*}(\tau -\tau' )\phi (\tau' ,\zeta )\phi ^{+} (\tau' ,\zeta )+
 \Gamma  ^{+ R}(\tau ,\zeta )\right]\phi ^{+} (\tau ,\zeta ),
\end{eqnarray}
The equations in \( \phi  \) and \( \phi^{+}  \) both have the same
additive noises and identical mean values, only differing by the independent
parts of the multiplicative noise sources - which therefore generate nonclassical
quantum statistics.

The causal linear response function \( g(\tau ) \) is defined as: 
\begin{eqnarray}
g(\tau )=\frac{\gamma (\tau t_{0})x_{0}}{v}\, . & 
\end{eqnarray}
If the Fourier transform of this function is \( \widetilde{g}(\Omega )=g(\Omega )+ig'(\Omega ) \),
then we can relate this to dimensionless intensity gain $\alpha ^{G}(\Omega )$ and loss $\alpha ^{A}(\Omega )$, at a relative
(dimensionless) detuning of \( \Omega  \) , by: 
\begin{equation}
2g(\Omega )=\alpha ^{A}(\Omega )-\alpha ^{G}(\Omega )\, \, .
\end{equation}

The causal nonlinear response function \( h(\tau ) \) is normalized so that
\( \int h(\tau )d\tau =1 \), and it includes both electronic and Raman nonlinearities:
\begin{eqnarray}
h(\tau )=h^{E}(\tau )+h^{R}(\tau )=\frac{\overline{n}x_{0}\chi (\tau t_{0})}{v^{2}}\, . & 
\end{eqnarray}
 The Raman response function \( h^{R}(\tau ) \) causes effects like the soliton
self-frequency shift\cite{s348}.  The response function Fourier transform is given by:
\begin{equation}
\widetilde{h}(\Omega )=\int dt\exp (i\Omega \tau )h(\tau )=h^{\prime }(\Omega )+ih^{\prime \prime }(\Omega ).
\end{equation}
This definition has the property that the value of $\widetilde{h}(\Omega )=\widetilde{h}(\omega t_0 )$ is a dimensionless number, which depends on the frequency $\omega$ only, independent of the time-scale used for normalization.  The Raman gain, whose spectrum has been extensively
measured\cite{e78}, can be modeled as a sum of \( n \) Lorentzians, as
explained in section \ref{RH} and as illustrated in Fig.~1. 

This expansion as \( n \) Lorentzians gives a response function of the form 
\begin{eqnarray}
h^{R}(t/t_{0})=\Theta (t)\sum _{j=0}^{n}F_{j}\delta _{j}t_{0}\exp(-\delta _{j}t)\sin (\omega _{j}t), & 
\end{eqnarray}
 It is most convenient to express these in terms of dimensionless parameters
\( \Omega _{j}=\omega _{j}t_{0} \) and \( \Delta _{j}=\delta _{j}t_{0} \),
giving: 
\begin{eqnarray}
h^{R}(\tau )=\Theta (\tau )\sum _{j=0}^{n}F_{j}\Delta _{j}\exp(-\Delta _{j}\tau )\sin (\Omega _{j}\tau ).
\end{eqnarray}
 Here \( \Delta _{j} \) are the equivalent dimensionless widths (corresponding
to damping), and the \( \Omega _{j} \) are the dimensionless center frequencies,
all in normalized units. It is useful to compare these results with the dimensionless
Raman gain \( \alpha ^{R}(\Omega ) \) , normalized following Gordon\cite{s348},
which uses a characteristic time-scale of \( t_{0} \) . The relationship of
macroscopic coupling \( R(\omega ) \) to measured Raman gain \( \alpha ^{R}(\Omega ) \)
is \( R^{2}(\omega )={\chi \alpha ^{R}(\omega t_{0})}/{2\pi } \). It follows
that the dimensionless gain function is 
\begin{equation}
\alpha ^{R}(\Omega )=2|h^{\prime \prime }(\Omega )|\, \, .
\end{equation}

These stochastic partial differential equations can be discretized and, without
any further approximation, can be numerically simulated\cite{s23,Werner1997}
using a split-step Fourier integration routine. The equations include all the
currently known noise physics significant in soliton propagation, including
effects like the soliton self-frequency shift. Guided acoustic wave Brillouin
scattering \cite{s795,s109,s80} noise sources are included in the Raman gain
function. These have little effect on the position of an isolated soliton, but
are important for long-range soliton collision effects\cite{s20} that occur
in pulse-trains.

\subsection{Initial conditions }

The initial conditions for the calculations could involve any required quantum
state, if the +$P$ representation is used. In the case of the Wigner equations,
only a subset of possible states can be represented with a positive probability
distribution. The usual initial condition is the multi-mode coherent state,
since this is the simplest model for the output of mode-locked lasers. In general,
there could be extra technical noise. We note that the choice of a coherent
state is the simplest known model of a laser sources. To represent this in the
positive-$P$ distribution is simple; one just takes 
\begin{equation}
\phi _{P}(\tau ,0)=[\phi _{P}^{+}(\tau ,0)]^{*}=\langle \widehat{\phi }(\tau ,0)\rangle.
\end{equation}
 In the Wigner case, which corresponds to symmetric operator ordering, one must
also include complex quantum vacuum fluctuations, in order to correctly represent
operator fields. For coherent inputs, the Wigner vacuum fluctuations are Gaussian,
and are correlated as 
\begin{eqnarray}
\langle \phi _{W}(\tau ,0)\rangle  & = & \langle \widehat{\phi }(\tau ,0)\rangle \nonumber  \\
\langle \Delta \phi _{W}(\tau ,0)\Delta \phi _{W}^{*}(\tau' ,0)\rangle  & = & \frac{1}{2\overline{n}}\delta (\tau -\tau' ).
\end{eqnarray}
 We note that these equations imply that an appropriate correction is made for
losses at the input interface, so that the mean-field evolution is known at
the fiber input face.

\subsection{Wigner noise}

Both fiber loss and the presence of a gain medium contribute quantum noise to
the equations in this symmetrically ordered representation. The complex gain/absorption
noise enters the Wigner equation through an additive stochastic term \( \Gamma  \),
whose correlations are obtained by averaging the normally and anti-normally
ordered reservoir correlation functions given previously, together with appropriate
variable changes. This symmetrically ordered noise source is present for both
gain and loss reservoirs. Thus, 
\begin{equation}
\label{gain_{c}or}
\langle \Gamma (\Omega ,\zeta )\Gamma ^{*}(\Omega' ,\zeta' )\rangle =\frac{(\alpha ^{G}(\Omega )+\alpha ^{A}(\Omega ))}{2\overline{n}}\delta (\zeta -\zeta' )\delta (\Omega -\Omega' ),
\end{equation}
 where \( \Gamma (\Omega ,\zeta ) \) is the Fourier transform of the noise
source: 
\begin{eqnarray}
\Gamma (\Omega ,\zeta ) & = & \frac{1}{\sqrt{2\pi }}\int ^{\infty }_{-\infty }d\tau \Gamma (\tau ,\zeta )\exp(i\Omega \tau )\nonumber \\
\Gamma ^{*}(\Omega ,\zeta ) & = & \frac{1}{\sqrt{2\pi }}\int ^{\infty }_{-\infty }d\tau \Gamma ^{*}(\tau ,\zeta )\exp(-i\Omega \tau ).
\end{eqnarray}
 Similarly, the real Raman noise, which appears as a multiplicative stochastic
variable \( \Gamma ^{R} \), has correlations 
\begin{equation}
\label{Raman_{c}or}
\langle \Gamma ^{R}(\Omega ,\zeta )\Gamma ^{R*}(\Omega' ,\zeta' )\rangle =\frac{\alpha ^{R}(|{\Omega }|)}{\overline{n}}\left[ n_{th}(|\Omega |/t_{0})+\frac{1}{2}\right] \delta (\zeta -\zeta' )\delta (\Omega -\Omega' )\, .
\end{equation}

Thus the Raman noise is strongly temperature-dependent, but it also contains
a spontaneous component which provides vacuum fluctuations even at \( T=0 \).
Since the spontaneous component can occur through coupling to either a gain
or a loss reservoir, in a symmetrically ordered representation, it is present
for both positive and negative frequency detunings.

It must be remembered here that the noise terms in the Wigner representation
do not correspond to normally ordered correlations, and so have no direct
interpretation in terms of photodetection experiments. Any predictions made
with this method of calculation need to be corrected by subtracting the appropriate
commutators, to convert the results into a normally ordered form. This is the
reason why there is no obvious distinction between the amplifier and absorber cases.

\subsection{+$P$ noise}

The positive P-representation is a useful alternative strategy, because it does
not require truncation of higher order derivatives in a Fokker-Planck equation,
and corresponds directly to observable normally ordered, time-ordered operator
correlations. It has no vacuum fluctuation terms. Provided the phase-space boundary
terms are negligible, one can then obtain a set of c-number stochastic differential
equations in a phase-space of double the usual classical dimensions. These are
very similar to the classical equations. Here the additive stochastic term is
as before, except it \textit{only} depends on the gain term \( \alpha ^{G} \);
the conjugate term \( \Gamma ^{*} \) is used in the \( \phi ^{+} \) equation:
\begin{eqnarray}
\langle \Gamma (\Omega ,\zeta )\, \Gamma ^{*}(\Omega ',\zeta ^{\prime })\rangle =\frac{\alpha ^{G}(\Omega )}{\overline{n}}\delta (\zeta -\zeta' )\delta (\Omega -\Omega' ) .
\end{eqnarray}
 Since this representation is normally ordered, the only noise sources present
are due to the gain reservoirs. There is no vacuum noise term for the absorbing
reservoirs, because absorption simply maps a coherent state into another coherent
state.

The complex terms \( \Gamma ^{R} \), \( \Gamma ^{R+} \) include both Raman
and electronic terms (through \( {h}'(\Omega ) \)). As elsewhere in this paper,
we regard \( \Gamma ^{R+}(\Omega ,\zeta ) \) as a hermitian conjugate Fourier
transform (with the opposite sign frequency exponent):
\begin{eqnarray}
\Gamma ^{+}(\Omega ,\zeta ) & = & \frac{1}{\sqrt{2\pi }}\int ^{\infty }_{-\infty }d\tau \Gamma ^{+}(\tau ,\zeta )\exp(-i\Omega \tau ).
\end{eqnarray}
This quantity is not the same as \( \Gamma ^{R*}(\Omega ,\zeta ) \), since it involves a noise source that is in general independent. In some cases, where classical noise is dominant (and nonclassical squeezing is negligible), we can ignore this fact, and approximately set \( \Gamma ^{R+}(\Omega ,\zeta ) =\Gamma ^{R*}(\Omega ,\zeta )  \). More generally, we obtain the following results:
\begin{eqnarray}
\langle \Gamma ^{R}(\Omega ,\zeta )\, \Gamma ^{R}(\Omega ',\zeta ')\rangle  & = & \delta (\zeta -\zeta ')\, \delta (\Omega +\Omega ')\left\{ \left[n_{th}(|\Omega |/t_{0})+1/2\right]\alpha ^{R}(|\Omega |)-i\, {h}'(\Omega )\right\} /{\overline{n}}\nonumber \\
\langle \Gamma ^{R+}(\Omega ',\zeta ')\, \Gamma ^{R}(\Omega ,\zeta )\rangle  & = & \delta (\zeta -\zeta ')\, \delta (\Omega -\Omega ')\left[n_{th}(|\Omega |/t_{0})+\Theta (-\Omega )\right]\alpha ^{R}(|\Omega |)/{\overline{n}}.
\end{eqnarray}
This equation is an expected result, since it states that when \( \Omega <0 \)
the spectral intensity of noise due to the Stokes process, in which a photon
is down-shifted in frequency by an amount \( \Omega  \) with the production
of a phonon of the same frequency, is proportional to \( n_{th}^{}+1. \) However
the anti-Stokes process in which a phonon is absorbed (\( \Omega >0 \)) is
only proportional to \( n_{th}^{}. \) Thus, at low temperatures the only direct
noise effect is that due to the Stokes process, which can be interpreted physically
as originating in spontaneous photon emission, detectable through photodetection.

As one might expect, the two forms of equation are identical at high phonon
occupation numbers, when classical noise is so large that it obscures the differences
due to the operator orderings of the two representations. Another, less obvious,
result is that the two equations have identical additive noise sources, provided
the gain and loss are balanced. To understand this, we can see that in the absence
of any net gain or loss, the differences in the operator correlations due to
ordering is a constant, contained in the initial conditions.

However, when gain and loss are not equal, the additive noise sources are quite different.
In particular, the Wigner representation has noise contributions from both
types of reservoir. On the other hand, the normally ordered +$P$ method only
leads to additive noise when there is a real fluorescent field present, which
is detectable through photodetection. This corresponds physically to some kind
of gain, either due to the presence of an amplifier, or through Raman effects.

In general, the Wigner and +$P$ reservoir correlations are obtainable simply by
examining the expectation values of the Heisenberg reservoir terms, with symmetric
and normal ordering respectively. The additional term proportional to \( {h}'(\Omega ) \)
in the +$P$ noise terms is due to dispersive nonlinear effects, and gives rise
to a nonclassical noise source which is responsible for the observed quantum
squeezing effects. Extensions required to treat polarization dependent Raman scattering are given elsewhere\cite{Drummond}.

\section{Conclusions}

\label{Cs} Our major conclusion is that quantum noise effects due to the intrinsic
finite-temperature phonon reservoirs and finite bandwidth amplification or absorption can be readily modeled using stochastic equations. The equations themselves
have the usual classical form, together with correction terms that we can describe as quantum noise terms. The precise form of the correction terms depends in detail
on the representation employed (although this difference is purely due to operator
ordering), as well as the physical origin of the reservoir couplings. These
correction terms can be non-Markovian or nonuniform in space. The generation
of the corresponding stochastic noises is a straightforward numerical procedure,
and generally much simpler than the use of noncommuting operators. By contrast,
the original operator equations have no practical numerical solution in most
cases, due to the exponential growth of the dimension of the underlying Hilbert
space with the number of modes and photons involved.

Detailed applications to short-pulse soliton communications will be given in
a subsequent paper\cite{DruCor99b}. In general, the increasing bandwidth, reducing pulse energies
and greater demands placed on fiber communications and sensors  mean that these
quantum limits are becoming increasingly important. Already, limits set by quantum
amplifiers are known to have great significance in long-distance laser-amplified
communications systems. We note that the quantum theory given here also establishes
the levels of quantum noise in silica fibers in more general situations. Examples
of this are for dispersion-managed fiber communications \cite{Smith1997,Lakoba1999},
and for fiber ring lasers with relatively low gain\cite{Collings1998}. Similarly,
these equations set the limits for experiments using spectral filtering and
related techniques to generate sub-shot-noise pulses\cite{Fri1996,Werner1999PRA}
in optical fibers.

\acknowledgments

We would like to acknowledge helpful comments on this paper by Wai S. Man.


\begin{thebibliography}{10}
\bibitem{s14} S.~J. Carter, P.~D. Drummond, M.~D. Reid, and R.~M. Shelby, ``Squeezing of
Quantum Solitons,'' Physical Review Letters \textbf{58}, 1841--1844 (1987). 
\bibitem{2} P.~D. Drummond and S.~J. Carter, ``Quantum-Field Theory of Squeezing in Solitons,''
Journal of the Optical Society of America B \textbf{4}, 1565--1573 (1987).
\bibitem{Solexp} M. Rosenbluh and R. M. Shelby, ``Squeezed Optical Solitons'', Phys. Rev.
Lett. \textbf{66}, 153--156 (1991); P.D. Drummond, R. M. Shelby, S. R. Friberg and Y. Yamamoto, ``Quantum solitons in optical fibres,'' Nature \textbf{365}, 307--313 (1993).
\bibitem{s35} J.~P. Gordon and H.~A. Haus, ``Random Walk of Coherently Amplified Solitons
in Optical Fiber Transmission,'' Optics Letters \textbf{11}, 665--667 (1986). 
\bibitem{s53} H.~A. Haus and W.~S. Wong, ``Solitons in Optical Communications,'' Reviews
of Modern Physics \textbf{68}, 423--444 (1996). 
\bibitem{DruCor99b}  P.~D. Drummond and J.~F. Corney, ``Quantum noise in optical fibers {II}: {R}aman jitter in soliton communications,''   {\em submitted to Journal of the Optical Society of America B}.  
\bibitem{s208}P.~D. Drummond, ``Electromagnetic quantization in dispersive inhomogeneous
nonlinear dielectrics,'' Physical Review A \textbf{42}, 6845--6857 (1990). 
\bibitem{s21} P.~D. Drummond and S.~J. Carter, ``Quantum-Field Theory of Squeezing in Solitons,''
Journal of the Optical Society of America B \textbf{4}, 1565--1573 (1987); P.~D.
Drummond, S.~J. Carter, and R.~M. Shelby, ``Time Dependence of Quantum Fluctuations
in Solitons,'' Optics Letters \textbf{14}, 373--375 (1989).
\bibitem{s90} B. Yurke and M.~J. Potasek, ``Solution to the Initial Value Problem for the
Quantum Nonlinear Schroedinger Equation,'' Journal of the Optical Society of
America B \textbf{6}, 1227--1238 (1989). 
\bibitem{Drumhill} P. D. Drummond and M. Hillery, ``Quantum theory of dispersive electromagnetic
modes'', Phys. Rev. A \textbf{59}, 691--707 (1999). 
\bibitem{PZL} E. Power, S. Zienau, ``Coulomb gauge in nonrelativistic quantum electrodynamics
and the shape of spectral lines'', Philos. Trans. Roy. Soc. Lond. A\textbf{251},
427 -454 (1959); R. Loudon, \textit{The Quantum Theory of Light} (Clarendon
Press, Oxford, 1983). 
\bibitem{Hillery} M. Hillery and L. D. Mlodinow, ``Quantization of electrodynamics in nonlinear
dielectric media'', Phys. Rev. A\textbf{30}, 1860 (1984). 
\bibitem{Bloembergen} N. Bloembergen, \emph{Nonlinear Optics}, (Benjamin, New York, 1965). 
\bibitem{Drummond} P. D. Drummond, ``Quantum Theory of Fiber-Optics and Solitons'',
in \emph{Coherence and Quantum Optics VII}, J. Eberly, L. Mandel and E. Wolf (Eds), 323--332 (Plenum Press, New York, 1996).
\bibitem{s05} G.~P. Agrawal, \emph{Nonlinear Fiber Optics}, 2nd ed. (Academic Press, 1995), pp 28--59. 

\bibitem{CD}S.~J. Carter and P.~D. Drummond, ``Squeezed Quantum Solitons and Raman Noise,''
Physical Review Letters \textbf{67}, 3757--3760 (1991). 
\bibitem{s55} F.~X. Kartner, D.~J. Dougherty, H.~A. Haus, and E.~P. Ippen, ``Raman Noise
and Soliton Squeezing,'' Journal of the Optical Society of America B \textbf{11},
1267--1276 (1994). 
\bibitem{s76} Y. Lai and S.-S. Yu, ``General Quantum Theory of Nonlinear Optical-Pulse Propagation,''
Physical Review A \textbf{51}, 817--829 (1995); S.-S. Yu and Y. Lai, ``Impacts
of Self-Raman Effect and Third-Order Dispersion on Pulse Squeezed State Generation
Using Optical Fibers,'' Journal of the Optical Society of America B \textbf{12},
2340--2346 (1995). 
\bibitem{s34}T. von Foerster and R.~J. Glauber, ``Quantum Theory of Light Propagation in
Amplifying Media,'' Physical Review A \textbf{3}, 1484--1511 (1971); I. A.
Walmsley and M. G. Raymer, ``Observation of Macroscopic Quantum Fluctuations in Stimulated {R}aman
Scattering,'' Phys. Rev. Lett. \textbf{50}, 962--965 (1983). 
\bibitem{Levenson}M. D. Levenson, \emph{Introduction to Nonlinear Laser Spectroscopy} (Academic
Press, New York, 1982). 
\bibitem{BellDean70}P. Dean, ``The Vibrational Properties of Disordered Systems: 
Numerical Studies'', Rev. Mod. Phys. \textbf{44}, 127 (1972). 
\bibitem{e78} R.~H. Stolen, C. Lee, and R.~K. Jain, ``Development of the Stimulated Raman
Spectrum in Single-Mode Silica Fibers,'' Journal of the Optical Society of
America B \textbf{1}, 652--657 (1984); D.~J. Dougherty, F.~X. Kartner, H.~A. Haus, and E.~P. Ippen, ``Measurement
of the Raman Gain Spectrum of Optical Fibers,'' Optics Letters \textbf{20},
31--33 (1995); R.~H. Stolen, J.~P. Gordon, W.~J. Tomlinson, and H.~A. Haus,
``Raman Response Function of Silica-Core Fibers,'' Journal of the Optical
Society of America B \textbf{6}, 1159--1166 (1989).  
\bibitem{s795} R.~M. Shelby, M.~D. Levenson, and P.~W. Bayer, ``Guided acoustic-wave Brillouin scattering'', Physical Review B \textbf{31}, 5244--5252 (1985).
\bibitem{s80} R.~M. Shelby, P.~D. Drummond, and S.~J. Carter, ``Phase-Noise Scaling in Quantum
Soliton Propagation,'' Physical Review A \textbf{42}, 2966--2796 (1990). 
\bibitem{s109} K. Bergman, H.~A. Haus, and M. Shirasaki, ``Analysis and Measurement of {GAWBS} Spectrum in a Nonlinear Fiber Ring,'' Applied Physics B \textbf{55}, 242--249
(1992). 
\bibitem{s20} K. Smith and L.~F. Mollenauer, ``Experimental observation of soliton interaction over long fiber paths: discovery of a long-range interaction,'' Optics Letters \textbf{14}, 1284--1286 (1989);  E.~M. Dianov, A.~V. Luchnikov, A.~N. Pilipetskii, and A.~M. Prokhorov, ``Long-Range
Interaction of Picosecond Solitons Through Excitation of Acoustic Waves in Optical
Fibers,'' Applied Physics B \textbf{54}, 175--180 (1992).  
\bibitem{Perlmutter} S. ~H. Perlmutter, M.~D. Levenson, R.~M. Shelby and M. ~B. Weissman, ``Inverse-Power-Law Light scattering in Fused-Silica Optical Fiber'', Phys. Rev. Lett. \textbf{61}, 1388--1391 (1988); ``Polarization properties of quasielastic light scattering in fused-silica optical fiber'', Physical Review B \textbf{42}, 5294--5305 (1990). 


\bibitem{Mears} R. J. Mears, L. Reekie, I. M. Jauncey, D. N. Payne, `` Low-noise erbium-doped fibre amplifier operating at $1.54\mu m$,'' Electron. Lett. 23, 1026--1028 (1987). 
\bibitem{Desurvire} E. Desurvire, \emph{Erbium-Doped Fiber Amplifiers, Principles and Applications}
(Wiley, New York, 1993). 
\bibitem{DrumRayn} P. D. Drummond and M. G. Raymer, ``Quantum theory of propagation of nonclassical
radiation in a near-resonant medium'' Physical Review A \textbf{44}, 2072--2085 (1991).

\bibitem{e36} L.~F. Mollenauer, ``Solitons in optical fibers and the soliton laser'', Philosophical Transactions of the Royal Society of London
A \textbf{315}, 435 (1985); L.~F. Mollenauer, R.~H. Stolen, and J.~P. Gordon, ``Experimental observation of picosecond pulse narrowing and solitons in optical fibers,'' Physical Review Letters \textbf{45}, 1095--1098 (1980). 
\bibitem{s348} J.~P. Gordon, ``Theory of the Soliton Self-Frequency Shift'', 
Optics Letters \textbf{11}, 662--664 (1986);
F. M. Mitschjke and L. F. Mollenauer, ``Discovery of the Soliton Self-Frequency Shift'', Optics Lett. \textbf{11}, 659--661 (1986). 
\bibitem{s23} P.~D. Drummond and A.~D. Hardman, ``Simulation of Quantum Effects in Raman-Active
Waveguides,'' Europhysics Letters \textbf{21}, 279--284 (1993); 
P.~D. Drummond and W. Man, ``Quantum Noise in Reversible Soliton Logic,''
Optics Communications \textbf{105}, 99--103 (1994).
\bibitem{s12} S.~J. Carter, ``Quantum Theory of Nonlinear Fiber Optics: Phase-Space representations,''
Physical Review A \textbf{51}, 3274--3301 (1995).
\bibitem{Werner1997} M. J. Werner and P. D. Drummond, ``Robust algorithms for solving stochastic
partial differential equations'', J. Comp. Phys. \textbf{132}, 312--326 (1997).

\bibitem{Smith1997} N. J. Smith, N. J. Doran, W. Forysiak, and F. M. Knox, ``Soliton transmission
using periodic dispersion compensation\char`\"{}, IEEE J. Lightwave Technol.,
\textbf{15}, 1808--1822 (1997). 
\bibitem{Lakoba1999} T. I. Lakoba, and D. J. Kaup, ``Influence of the Raman effect on dispersion-managed
solitons and their interchannel collisions'', Optics Letts. \textbf{24}, 808-810 (1999).
\bibitem{Collings1998}S. Namiki, C. X. Yu, and H. A. Haus, ``Observation of nearly quantum-limited
timing jitter in an all-fiber ring laser\char`\"{}, Journal of the Optical Society of America {B} \textbf{13},
2817--2823 (1996); B. C. Collings, K. Bergman, and W. H. Knox, ``Stable multigigahertz pulse-train
formation in a short-cavity passively harmonic mode-locked erbium/ytterbium
fiber laser\char`\"{}, Opt. Letts. \textbf{23}, 123--125 (1998). 
\bibitem{Fri1996} S. R. Friberg, S. Machida, M. J. Werner, A. Levanon and T. Mukai, ``Observation of Optical Soliton Photon-Number Squeezing,'' Phys. Rev.
Lett. \textbf{77}, 3775--3778 (1996); S. Spalter, M. Burk, U. Strossner, M. Bohm,
A. Sizmann, and G. Leuchs, ``Photon number squeezing of spectrally filtered sub-picosecond optical solitons,'' Europhys. Lett. \textbf{38}, 335--340 (1997); D. Krylov
and K. Bergman, ``Amplitude-squeezed solitons from an asymmetric fiber interferometer\char`\"{},
Opt. Letts. \textbf{23}, 1390--1392, (1998). 
\bibitem{Werner1999PRA} M. J. Werner, ``Raman-induced photon correlations in optical fiber solitons'',
Phys. Rev. A \textbf{60}, R781--R784 (1999). 
\begin{figure}

\caption{The parallel polarization Raman gain \protect\protect\protect\( |\Im \{\tilde{h}(\omega t_0 )\}| = |h''(\omega t_0 )|\protect \protect \protect \)
for the 11-Lorentzian model (continuous lines) and the single-Lorentzian model
(dashed lines), for a temperature of \protect\protect\protect\( T=300K\protect \protect \protect \). }
\end{figure}
\begin{table}

\caption{Fitting parameters for the 11-Lorentzian model of the Raman gain function \protect\protect\protect\( h^{R}(t/t_{0})\protect \protect \protect \).
All frequencies are in T.rad/s.}
\begin{tabular}{|c|c|c|c|}
\hline 
\( j \)&
 \( F_{j} \)&
 \( \omega _{j} \)&
 \( \delta _{j} \)\\
\hline 
0 &
 0.16 &
 0.005 &
 0.005 \\
\hline 
1 &
 -0.3545 &
 0.3341 &
 8.0078 \\
\hline 
2 &
 1.2874 &
 26.1129 &
 46.6540 \\
\hline 
3 &
 -1.4763 &
 32.7138 &
 33.0592 \\
\hline 
4 &
 1.0422 &
 40.4917 &
 30.2293 \\
\hline 
5 &
 -0.4520 &
 45.4704 &
 23.6997 \\
\hline 
6 &
 0.1623 &
 93.0111 &
 2.1382 \\
\hline 
7 &
 1.3446 &
 99.1746 &
 26.7883 \\
\hline 
8 &
 -0.8401 &
 100.274 &
 13.8984 \\
\hline 
9 &
 -0.5613 &
 114.6250 &
 33.9373 \\
\hline 
10 &
 0.0906 &
 151.4672 &
 8.3649  \\
\hline 
\end{tabular}
\label{lorfit}
\end{table}

\end{thebibliography}
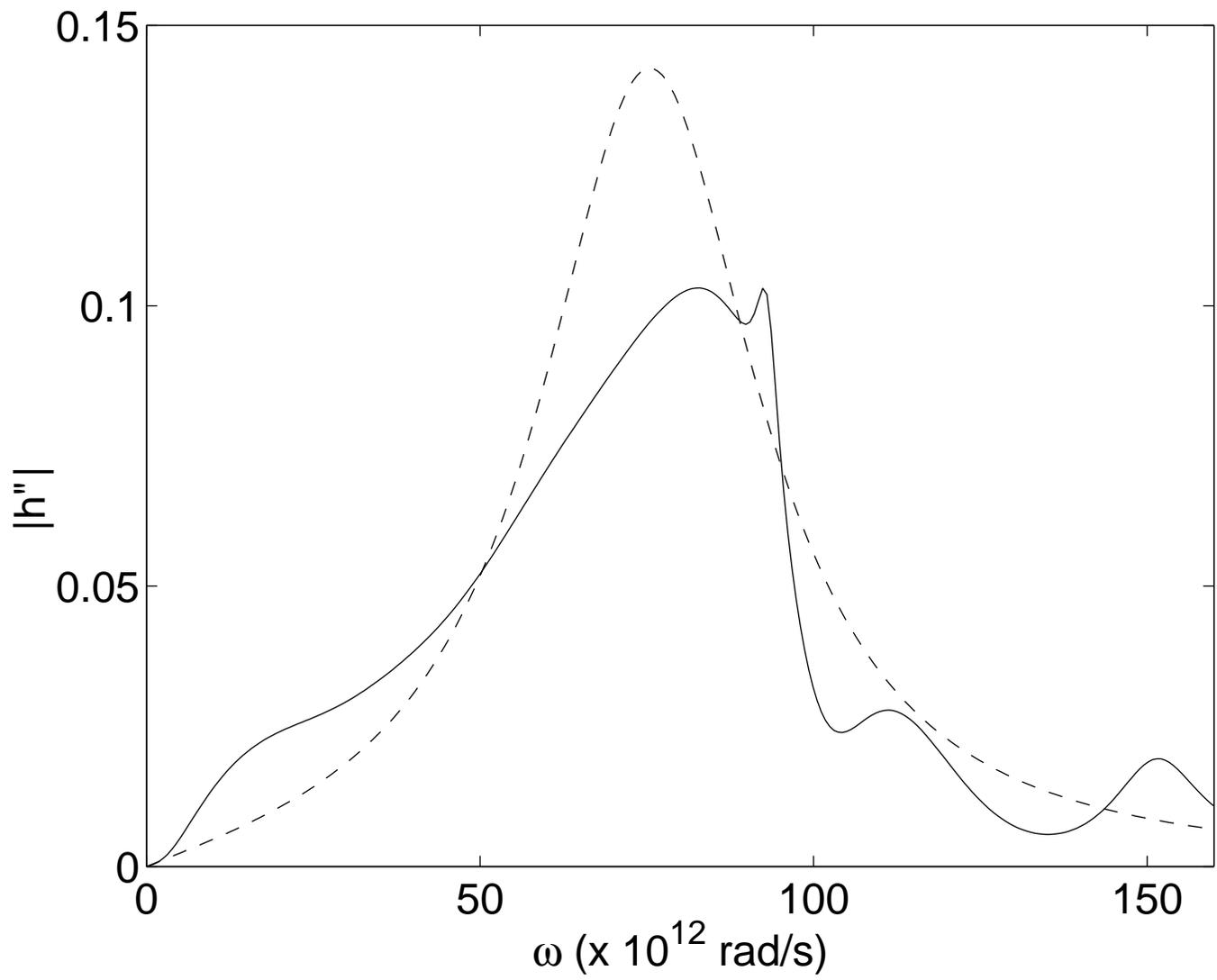
\end{document}